# Security Scheme for Distributed DoS in Mobile Ad Hoc Networks

Sugata Sanyal[1], Ajith Abraham[2], Dhaval Gada[3], Rajat Gogri[3], Punit Rathod[3], Zalak Dedhia[3] and Nirali Mody[3]

[1]School of Technology and Computer Science, Tata Institute of Fundamental Research, India, `sanyal@tifr.res.in`
[2]Computer Science Department, `Oklahoma State University, USA`
`ajith.abraham@ieee.org`
[3]Mumbai University, India

**Abstract.** In Mobile Ad hoc Networks (MANET), various types of Denial of Service Attacks (DoS) are possible because of the inherent limitations of its routing protocols. Considering the Ad hoc On Demand Vector (AODV) routing protocol as the base protocol it is possible to find a suitable solution to overcome the attack of initiating / forwarding fake Route Requests (RREQs) that lead to hogging of network resources and hence denial of service to genuine nodes. In this paper, a proactive scheme is proposed that could prevent a specific kind of DoS attack and identify the misbehaving node. Since the proposed scheme is distributed in nature it has the capability to prevent Distributed DoS (DDoS) as well. The performance of the proposed algorithm in a series of simulations reveal that the proposed scheme provides a better solution than existing approaches with no extra overhead.

## 1  Introduction

In an ad hoc wireless network where wired infrastructures are not feasible, energy and bandwidth conversation are the two key elements presenting research challenges. Limited bandwidth makes a network easily congested by control signals of the routing protocol. Routing schemes developed for wired networks seldom consider restrictions of this type. Instead, they assume that the network is mostly stable and the overhead for routing messages is negligible. Considering these differences between wired and wireless network, it is necessary to develop a wireless routing protocol that restricts congestion in the network [1][2][3][4][5][6].

This paper proposes minor modifications to the existing AODV routing protocol (RFC 3561) in order to restrict congestion in the network during a particular type of DoS attack. In addition to this it incurs absolutely no extra overhead [7]. The rest of this paper is organized as follows. In section 2, we describe the DoS attack caused due to RREQ flooding and its implications on the existing AODV driven MANET [8][9]. To combat this DoS attack a proactive [10] scheme is proposed in section 3. Section 4 presents an illustration to describe the implications of RREQ flooding on pure AODV and the modified AODV. To quantify the effectiveness of the proposed scheme, a DoS [11] attack was simulated in the mobile environment and its performance results are reported in section 5. Finally section 6 gives the conclusion and further work.

## 2. DoS Attack Due to RREQ Flooding

In AODV, a malicious node can override the restriction put by *RREQ_RATELIMIT* [12] (limit of initiating / forwarding RREQs) by increasing it or disabling it. A node can do so because of its self-control over its parameters. The default value for the *RREQ_RATELIMIT* is 10 as proposed by RFC 3561. A compromised node may choose to set the value of parameter *RREQ_RATELIMIT* to a very high number. This allows it to flood the network with fake RREQs [12] and lead to a kind of DoS attack. In this type of DoS attack a non-malicious node cannot fairly serve other nodes due to the network-load imposed by the fake RREQs. This leads to the following problems:

- Wastage of bandwidth
- Wastage of nodes' processing time (more overhead)
- Exhaustion of the network resources like memory (routing table entries)
- Exhaustion of the node's battery power

This further results in degraded throughput. Most of the network resources are wasted in trying to generate routes to destinations that do not exist or routes that are not going to be used for any communication. This implies that the existing version of AODV is vulnerable to such type of malicious behavior from an internal node (which is then termed as a compromised node).

## 3. Proposed Scheme

### 3.1. Overview

As mentioned earlier, the default value for *RREQ_RATELIMIT* is 10 RREQs/sec. This means each node is expected to observe some self-control on the number of RREQs it sends in one sec. A compromised node may choose to set the value of parameter *RREQ_RATELIMIT* to a very high number or even disable this limiting feature, thus allowing it to send large number of RREQ packets per second. The proposed scheme shifts the responsibility to monitor this parameter on the node's neighbor, thus ensuring the compliance of this restriction. This solves all of the problems (mentioned in section 2) caused due to flooding of RREQs from a compromised node. Thus instead of self-control, the control exercised by a node's neighbor results in preventing the flooding of RREQs.

### 3.2. RREQ_ACCEPT_LIMIT and RREQ_BLACKLIST_LIMIT

The proposal is based on the application of two parameters: *RREQ_ACCEPT_LIMIT* and *RREQ_BLACKLIST_LIMIT*. *RREQ_ACCEPT_LIMIT* denotes the number of RREQs that can be accepted and processed per unit time by a node. The purpose of this parameter is to specify a value that ensures uniform usage of a node's resources by its neighbors. RREQs exceeding this limit are dropped, but their timestamps are recorded. This information will aid in monitoring the neighbor's activities. In the simulations carried out, the value of this parameter was kept as three (i.e. three RREQs can be accepted per unit time). This value can be made adaptive, depending upon node metrics such as it memory, processing power, battery, etc.

The *RREQ_BLACKLIST_LIMIT* parameter is used to specify a value that aids in determining whether a node is acting malicious or not. To do so, the number of RREQs originated/forwarded by a neighboring node per unit time is tracked. If this count exceeds the value of *RREQ_BLACKLIST_LIMIT*, one can safely assume that the corresponding neighboring node is trying to flood the network with possibly fake RREQs. On identifying a neighboring node as malicious, it will be blacklisted. This will prevent further flooding of the fake RREQs in the network. The blacklisted node is ignored for a period of time given by *BLACKLIST_TIMEOUT* after which it is unblocked. The proposed scheme has the ability to block a node till *BLACKLIST_TIMEOUT* period on an incremental basis. The *BLACKLIST_TIMEOUT* period is doubled each time the node repeats its malicious behavior.

In the simulations the value of *RREQ_BLACKLIST_LIMIT* is kept as 10 (i.e. more than 10 RREQs per unit time results in flooding activity). By blacklisting a malicious node, all neighbors of the malicious node restrict the RREQ flooding. Also the malicious node is isolated due to this distributed defense and so cannot hog its neighbor's resources. The neighboring nodes are therefore free to entertain the RREQs from other genuine nodes. Nodes that are confident about the malicious nature of a particular node, can avoid using it for subsequent network functions. In this way genuine nodes are saved from experiencing the DoS attack.

### 3.3. Advantages of the Proposed Scheme

- The proposed scheme incurs no extra overhead, as it makes minimal modifications to the existing data structures and functions related to blacklisting a node in the existing version of pure AODV (RFC 3561).
- Also the proposed scheme is more efficient in terms of its resultant routes established, resource reservations and its computational complexity.
- If more than one malicious node collaborate, they in turn will be restricted and isolated by their neighbors, since they monitor and exercise control over forwarding RREQs by nodes. Thus the scheme successfully prevents DDoS attacks.

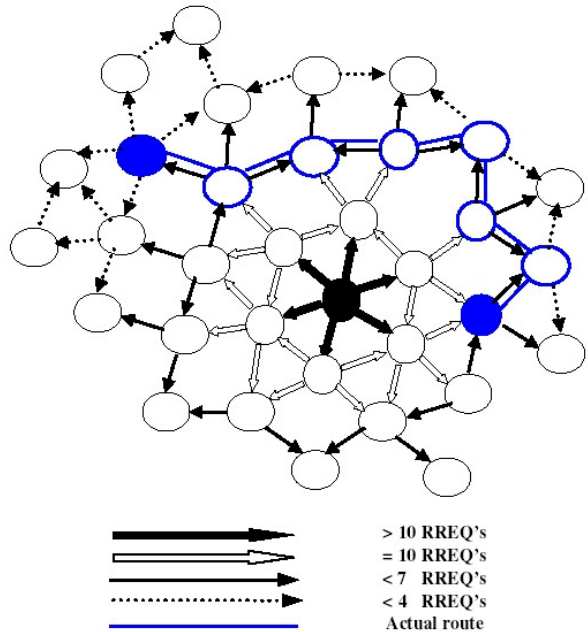

Figure 1. Illustration in original AODV

## 4. Algorithm Illustration

Figure 1 depicts the working in pure AODV routing protocol when an internal malicious node launches a DoS attack by flooding the network with RREQs. The black node depicts the malicious node and the blue nodes depict two genuine nodes that want to communicate with each other. The optimal route consists of four intermediate nodes including the malicious node and three of its neighbors. The malicious node floods the network by generating 10 RREQs per second as shown. Its immediate neighbors, (who are not mali-cious) observe the *RREQ_RATELIMIT* and hence each forward 10 RREQs only. Since at max three RREQs will be accepted from these nodes within one second, the neighbors of these nodes need to forward < seven RREQs and their neighbors in turn need to forward < four RREQs, as shown. Since the resources of the malicious node's neighbors are completely occupied in processing and forwarding the RREQ's originating from it, the route between the blue nodes, if it is established, will consist of greater number of intermediate nodes. Thus in effect a DoS attack is launched as the genuine nodes are deprived of the services of nodes whose resources are wasted due to flooding.

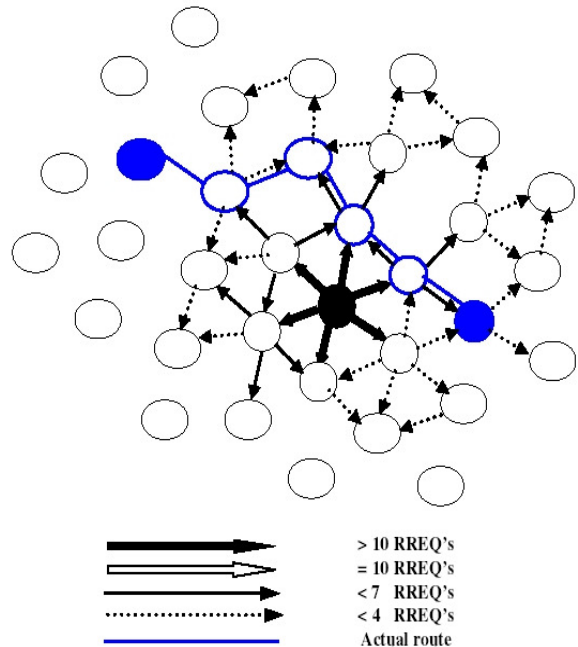

**Figure 2.** Illustration of the proposed AODV

Figure 2 illustrates the working procedure in the proposed AODV scheme. As shown in the figure, malicious node (depicted by the black node) floods RREQs in the network and two genuine nodes (depicted by blue nodes) want to communicate with each other. In this scheme, the no. of RREQs that can be accepted from a neighbor is limited. Hence, the neighbors of the malicious node, will only accept and forward three RREQ packets received from it within a time interval of one sec. This rate limit of three packets is to ensure fair share of a node's resources to all the neighbors. Moreover, whenever the malicious node crosses the RREQ_BLACKLIST_LIMIT of 10 RREQ packets within a time interval of one sec, its neighbors will blacklist it. Thus, in addition to limiting the clogging up of resources in the network, the proposed scheme also, isolates the malicious node. The route established in this scheme is expected to be the optimum route, which consists of minimum number of intermediate nodes. Thus, no DoS attack is experienced in the developed scheme.

## 5. Simulation/Experiments and Analysis

NS-2 simulator is used [13][14] for the implementation of the proposed scheme. The IEEE 802.11 [15] protocol is used for the MAC layer. The AODV protocol incorporated in NS-2 by Uppsala University, Sweden, was used as the base protocol. Modifications were made to this version of AODV protocol that confirms to RFC 3561. TCP was used as the transport protocol Radio transmission range is set as 250 meters. Traffic sources used are Constant-Bit-Rate (CBR) and the field configuration is 2000 x 2000m with 69 nodes.

### 5.1. Traffic Scenario

Node 0 is configured as the malicious node. It starts flooding the network with fake RREQ's at simulated time of one sec till time 17 secs. The traffic was generated such that the source and destination pairs are randomly spread over the entire network. The other source-destination pairs are shown in Table 1.

**Table 1**. Traffic generation summary

| Source  | Destination | Simulation time |
|---------|-------------|-----------------|
| Node 48 | Node 20     | 11-16 sec       |
| Node 18 | Node 27     | 5-12 sec        |
| Node 31 | Node 66     | 6-11 sec        |
| Node 45 | Node 16     | 9-12 sec        |

The performance evaluation of the proposed detection scheme involves study of two different aspects:
- Performance of original AODV protocol in presence of compromised nodes.
- Performance of proposed AODV protocol in presence of compromised nodes.

Each simulation was carried out for 17.2 seconds. The results for both cases have been observed. The following section gives the parameters that were measured for both the original and the modified protocols.

**5.2. Network Simulation Metrics**

The metrics are the important determinants of network performance, which have been used to compare the performance of the proposed scheme in the network with the performance of the original protocol. This study has been done to show that the proposed scheme enhances the security of the routing protocol without causing substantial degradation in network performance.

1) **End-to-End Delay:** Average time difference (in seconds) between the time of the packet receipt at the destination node, and the packet sending time at the source node.

2) **Round Trip Time (RTT):** Time difference between the receipt of the acknowledgement from the destination node to the source node, and the time of sending of the original packet at the source node.

3) **Average simulation processing time at nodes for a packet:** Time difference between the packet forwarding time and the packet receipt time at a given node.

4) **Average number of nodes receiving packets:** Sum of numbers of all the intermediate nodes (nodes between source and destination nodes) receiving packets sent by all the source nodes / number of received packets at all the destination nodes.

5) **Average number of nodes forwarding packets:** Sum of numbers of all the intermediate nodes (nodes between source and destination nodes) forwarding packets sent by all the source nodes / number of received packets at all the destination nodes.

6) **Delays between current and other node:** Shows end-to-end delays (in seconds) between current node (sender) and other node (receiver)

7) **Number of data packets dropped:** The number of data packets dropped at any given node. This is an important parameter because if the number of dropped packets increases, the throughput would decrease.

8) **Throughput:** It is sum of sizes (bits), or number (packets) of generated/sent/forwarded/received packets, calculated at every time interval and divided by its length. Throughput (bits) is shown in bits. Throughput (packets) shows numbers of packets in every time interval. Time interval length is equal to one second by default.

### 5.3. Performance Evaluation

This section consists of the results for the test cases. The recorded values are obtained by averaging over three runs for each test case.

**Acknowledgement packet receive time v/s RTT**

As simulation time increases the network resources available to the nodes vary. The availability of network resources is one of the parameter, which helps in deciding the RTT. Figure 3 shows the Graph of Acknowledgment Packet receive time versus RTT.

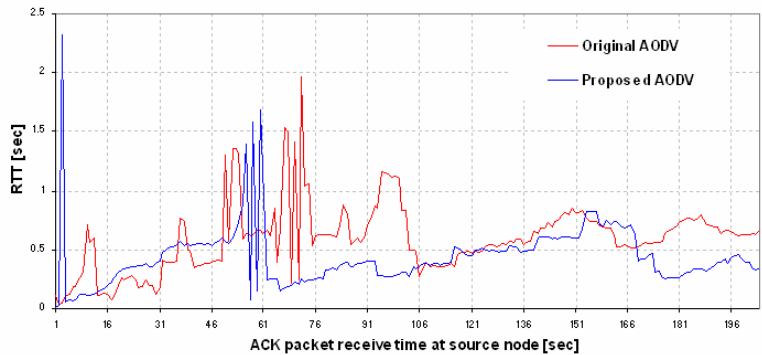

**Figure 3**: Acknowledgement packet receive time versus round trip time

It is evident from Figure 3 that as time proceeds; RTT is lesser in the proposed AODV scheme, as compared to the original scheme. This is because of the limit imposed on the number of RREQ packets being flooded in the network by malicious node and less number of intermediate nodes in the routes between genuine nodes.

**Dropped Packet Sum:**

The number of Packets dropped at a given instance of time in the simulation run determines the efficiency of the protocol. Figure 4, accommodates the information regarding the number of dropped packets throughout the simulation.

From figure 4, it is found that overall, the number of packets dropped using the proposed scheme is lesser than the number of packets dropped when using the original scheme. In the initial stages, the large amount of drops in the original scheme is due to the fact that the Flooding of RREQ in the network causes congestion, and the route formation for genuine requests is delayed. Thus, the buffered data packets are timed out and dropped. During the later stages the unavailability of network resources causes the data packets to be dropped. The improvement in the proposed scheme is due to the fact that there exists optimum utilization of the network resources and there is no overload, leading to comparatively lesser packet drops.

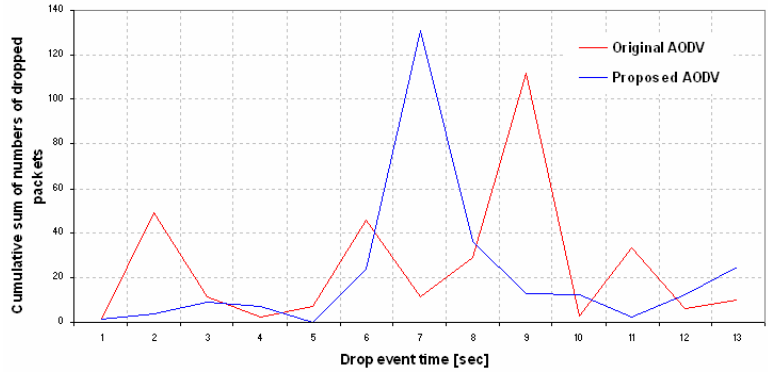

**Figure 4:** Throughput of dropping packets

### End-to-End Delay v/s Packet Size

Figure 5 depicts how the proposed method affects the end-to-end delay. This is the average delay of all data packets.

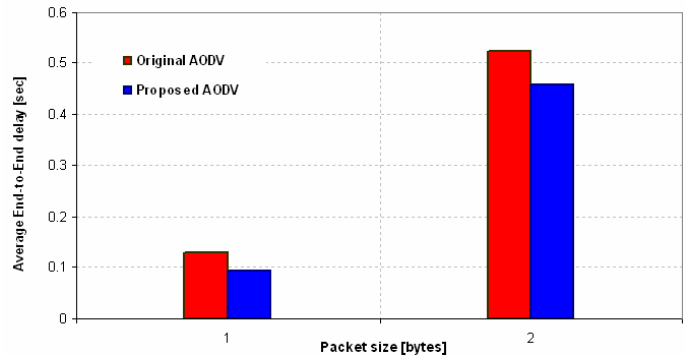

**Figure 5:** Packet size versus average end-to-end delay

The delay in case of both data and AODV packets in case of the proposed scheme is lesser compared to the original AODV

### Throughput of generating packets at an intermediate node in the route

The following graph shows throughput of generating packets at an intermediate node (numbered 48 in the sample simulation scenario) vs simulation time in seconds. The graph reflects the simulation time for which an intermediate node in the route generated packets in original AODV as compared to proposed AODV. In other words, it depicts how long the route through the intermediate node was valid during simulation.

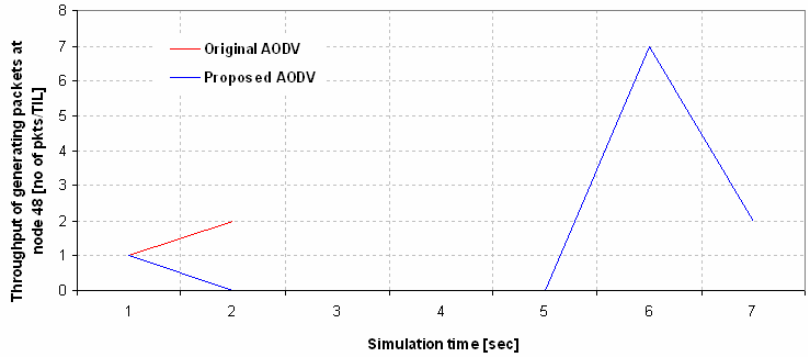

**Figure 6.** Throughput of generating packets at node 48

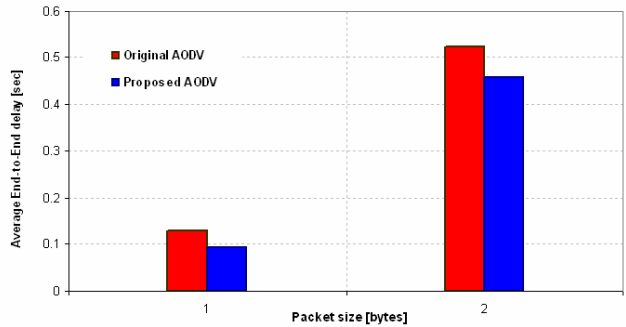

**Figure 7.** Average simulation processing time

In the scenario with original AODV protocol, the routes become invalid quickly when no replies (ACKs) for data packets are received due to clogging of network resources and the DoS attack. This is reflected in the graph by having > 0 throughput of generating packets for only two seconds of simulation time, after which, the route through the intermediate node 48 becomes invalid and hence resulting in 0 throughput of generating packets. However, in the scenario with the proposed AODV, the route through node 48 remains valid for longer period of simulation time and hence it has > 0 throughput of generating packets till simulation time 7 seconds (shown by blue line in the graph). Thus, it can be inferred from the graph that routes remain valid for longer periods of time under the proposed scheme.

**Packet size versus simulation time**

The comparison of simulation processing times as illustrated in Figure 6 reveals that the proposed scheme incurs no additional overhead as compared to the original scheme.

**Table 2.** Overall network simulation results

|  | Original AODV | Proposed AODV |
|---|---|---|
| Average End-to-end delay [sec] | 0.32539 | 0.27576 |
| Receiving packets | 0.4356328083 | 0.3580786026 |
| Forwarding packets | 0.4285714286 | 0.3499688085 |
| Average RTT | 0.58819 | 0.45346 |

**Network information for sample scenario**

Table 2 gives the comparative study of network information for original AODV and proposed AODV.

## 6. Conclusions and Future Work

The DoS attack caused due to RREQ flooding in ad hoc network can be successfully detected in the proposed scheme. The scheme can accurately detect the malicious nodes in the network. The malicious nodes identified are blacklisted and none of the genuine nodes in the network are wrongly accused of misbehaving. In the proposed scheme, there is an enhancement in the performance of the network in presence of compromised nodes.

Mobile computing and communication is an upcoming field, which is capturing the imagination of all the researchers worldwide. Thus the scope of enhancements and improvements is enormous. An immediate enhancement can be making the limit-parameters adaptive in nature. This can be done by making calculations based on parameters like memory, processing capability, battery power, and average number of requests per second in the network and so on. Further, the protocol can be made secure against other types of possible DoS attacks that threaten it.

NOTE:
AODV_modifications.zip
<http://www.tifr.res.in/~sanyal/papers/AODV_modifications.zip> contains all associated codes and its possible explanation. This has been added to provide some help to students and researchers who could interested in implementation issues. Please note that this is added with the best effort basis. On-line email help is not easily possible to provide.